\documentclass[fleqn,usenatbib]{mnras}

\usepackage{newtxtext,newtxmath}
\usepackage[T1]{fontenc}

\DeclareRobustCommand{\VAN}[3]{#2}
\let\VANthebibliography\thebibliography
\def\thebibliography{\DeclareRobustCommand{\VAN}[3]{##3}\VANthebibliography}

\usepackage{graphicx}	% Including figure files
\usepackage{amsmath}	% Advanced maths commands
\newcommand{\red}{}
\title[Reference turbulence profiles for SCAO/XAO]{Reference Optical Turbulence and Wind Profiles for Single Conjugate and Extreme Adaptive Optics}

\author[O. J. D. Farley]{
O. J. D. Farley$^{1}$\thanks{E-mail: o.j.d.farley@durham.ac.uk}
\\
$^{1}$Centre for Advanced Instrumentation, Department of Physics, Durham University, DH1 3LE, UK\\
}

\date{Accepted XXX. Received YYY; in original form ZZZ}

\pubyear{2022}

\begin{document}
\label{firstpage}
\pagerange{\pageref{firstpage}--\pageref{lastpage}}
\maketitle

\begin{abstract}
We present a simple method of extracting a small number of reference optical turbulence and wind profiles from a large dataset for single conjugate and extreme adaptive optics simulations. These reference profiles can be used in slow end-to-end adaptive optics simulations to represent the variability of the atmosphere. The method is based on the assumption that performance for these systems is correlated with integrated atmospheric parameters $r_0$, $\theta_0$ and $\tau_0$. Profiles are selected from a large dataset that conform concurrently to the distributions of these parameters, and hence represent the variability of the atmosphere as seen by the AO system. We also extend the equivalent layers method of profile compression to include wind profiles. The method is applied to stereo-SCIDAR data from ESO Paranal to extract five turbulence and wind profiles that cover a broad range in atmospheric variability, and we show using analytical AO simulation that this correlates to the equivalent range of AO-corrected Strehl ratios.
\end{abstract}

% Include between one and six keywords.
\begin{keywords}
adaptive optics -- site testing -- data analysis
\end{keywords}

%%%%%%%%%%%%%%%%%%%%%%%%%%%%%%%%%%%%%%%%%%%%%%%%%%

%%%%%%%%%%%%%%%%% BODY OF PAPER %%%%%%%%%%%%%%%%%%

\section{Introduction}
Adaptive optics (AO) techniques allow for partial compensation of the phase aberrations suffered by light propagating through atmospheric optical turbulence \citep{Davies2012}. In optical astronomy and increasingly other applications such as ground-space optical communications, we are concerned with the vertical propagation regime, in which plane-wave light propagates downwards through turbulence to the ground. The vertical distribution (profile) of the turbulence is one of the key determinants of the performance of AO systems. The profile is described by the refractive index structure constant $C_n^2 (h)$ as a function of height $h$ above the observer. The wind speed $v(h)$ and direction $\theta(h)$ associated with the turbulent layers is also of interest, since they define the temporal characteristics of the turbulence. 

The profile is primarily a meteorological phenomenon, and varies both spatially across different sites and temporally, changing on timescales as short as minutes up to seasons. Therefore for any particular location, characterisation of the profile requires long term measurement campaigns \citep[e.g.][]{Osborn2018a, Travouillon2009, Vernin2011} or modelling based on meteorological data \cite[e.g.][]{Osborn2018b, Masciadri2013}. These produce large datasets containing thousands of profiles.  

A common use of the profile is as input to Monte Carlo AO simulations, where they are used to generate and translate random phase screens which serve as a model of the turbulent atmosphere. However, these simulations are computationally intensive and hence large datasets with thousands of profiles cannot be used to fully characterise the AO performance. 

Reference profiles that reflect the variability in the underlying dataset must therefore be extracted for use in simulation \citep[e.g.][]{Sarazin2013, Sarazin2017}. For more complex multi-guide star tomographic AO systems, where some of the residual wavefront error terms such as tomographic error are a non-trivial function of the profile, this has been directly addressed via cluster analysis \citep{Farley2018} and using fast analytical AO simulation \citep{Farley2019}. An important conclusion of these works was that the common practice of averaging $C_n^2(h)$ in each altitude bin to produce a "typical" or "median" profile may actually produce profiles that do not have the expected characteristics and therefore produce erroneous results in simulation. It is better to select single profiles from the dataset that conform to the required characteristics for a particular system.

For single conjugate AO (SCAO), and its subset extreme AO (XAO), reference profiles are in principle easier to define since the system measures and corrects integrated turbulence along a single line of sight. This means the residual error terms are functions of integrated atmospheric parameters. However, there is not currently a procedure in the literature for extracting single profiles that conform to certain integrated parameters, without the averaging procedure previously described.

Here, a simple method of extracting reference profiles from a large dataset is presented that selects single profiles that conform to required distributions of integrated atmospheric parameters. We also present an extension of the equivalent layers method of profile compression \citep{Fusco2001} that accounts for the wind profile and guarantees conservation of said atmospheric parameters when the profile is compressed to a small number of layers suitable for use in Monte Carlo simulation.

We apply this method to the 2019A data release from the Stereo-SCIDAR at ESO Paranal, Chile, selecting five profiles from 19358 that represent 5th, 25th, 50th, 75th and 95th percentile in terms of the distribution of atmospheric parameters and also in terms of SCAO/XAO performance.

\section{Method}
The residual wavefront error after SCAO/XAO correction can be defined in terms of classical error terms
\begin{equation}
    \sigma_\phi^2 = \sigma_\mathrm{fitting}^2 + \sigma_\mathrm{aniso}^2 + \sigma_\mathrm{servo}^2 + \sigma_\mathrm{alias}^2 + \sigma_\mathrm{noise}^2 ,
\end{equation}
where $\sigma_\mathrm{fitting}^2$ is deformable mirror (DM) fitting error, $\sigma_\mathrm{aniso}^2$ angular anisoplanatic error, $\sigma_\mathrm{servo}^2$ temporal servo lag error, $\sigma_\mathrm{alias}^2$ wavefront sensor (WFS) aliasing and $\sigma_\mathrm{noise}^2$ WFS noise. From classical AO theory, $\sigma_\mathrm{fitting}^2$, $\sigma_\mathrm{aniso}^2$, $\sigma_\mathrm{servo}^2$, and $\sigma_\mathrm{alias}^2$ can be computed from atmospheric parameters $r_0$, $\theta_0$ and $\tau_0$, describing turbulence strength, angular correlation and temporal correlation respectively \citep{Rigaut1998, Fried1982, Fried1990}. It should be noted that these computations are based on a simplified AO model, and for a real system the error terms may only be correlated with atmospheric parameters as opposed to being defined by them. Our primary assumption is that this correlation is strong enough that the atmospheric parameters may be used as a proxy for AO performance.

Any reference turbulence profiles drawn from a dataset should therefore reflect the respective distributions of $r_0$, $\theta_0$ and $\tau_0$ if they are to reflect the variability in total wavefront error that arises from diverse turbulence conditions.

To accomplish this, we first compute the distributions of the parameters from the profiles, according to the definitions
\begin{equation}
    r_0 = \left( 0.423 k^2 \int_0^\infty C_n^2(h) \, \mathrm{d}h\right)^{-3/5}
\end{equation}
\begin{equation}
    \theta_0 = \left( 2.91 k^2 \int_0^\infty C_n^2(h) h^{5/3} \,\mathrm{d}h \right)^{-3/5}
\end{equation}
\begin{equation}
    \tau_0 = \left( 2.91 k^2 \int_0^\infty C_n^2(h) v(h)^{5/3} \, \mathrm{d}h \right)^{-3/5}
\end{equation}
with $k=2\pi/\lambda$ the wavevector for light of wavelength $\lambda$, $h$ the altitude, $C_n^2(h)$ the refractive index structure constant profile and $v(h)$ the wind velocity profile \citep{Fried1966,Fried1976,Roddier1981}.

Having obtained the distributions, for each value in the database we compute its quantile within that distribution 
\begin{equation}
    \label{eq:quantiles}
    Q(x) = \mathrm{Pr}(x>X) = 1 - F_X(x)
\end{equation}
where the quantile $Q$ of a parameter value $x$ in a distribution of values $X$ is the probability that $x$ is greater than each value in $X$, or equivalently the complement of the cumulative distribution function (CDF) of $X$ evaluated at $x$. This is therefore a normalisation of each parameter set to values between 0 and 1, with 0.5 indicating the median and so on. This is performed to ensure each parameter is treated with equivalent weight in the following analysis, whilst allowing arbitrary distributions of the parameters themselves.

The profiles are selected using a k-d tree search algorithm \citep{Maneewongvatana2002}, which allows rapid computation of nearest neighbours in an N-dimensional parameter space. In our case, we have a three dimensional space consisting of the quantiles of $r_0$, $\theta_0$ and $\tau_0$. We then search this space for data points that lie close to quantile values of all three distributions concurrently. For example, a search around the point (0.5,0.5,0.5) will return the profile that lies closest to the median values of all parameters. The choice of the quantile values may depend on the application, here we will use the values [0.05, 0.25, 0.5, 0.75, 0.95], corresponding to the 5th percentile, lower quartile, median, upper quartile and 95th pecentile of each distribution. This gives five profiles covering a broad range of variability in each parameter, and therefore a broad range of variability in AO performance. Provided the dataset is large enough, single profiles should be found that conform to these constraints within a small error. \red{The size of this error, which we can quantify as the euclidean distance in the 3-dimensional quantile space between the target and selected profile, is returned as part of the k-d tree search and can be used to assess the quality of the profiles selected. We will denote this quantity $\Delta Q$.}

If desired, additional parameters can be included in this search process. For example, scintillation index or Rytov variance may be useful to include if the AO system is required to work at low elevation angles. This is a trivial extension of the method, however it should be noted that a higher dimensional parameter space search makes it more difficult to select profiles that adhere to target values of all parameter distributions concurrently due to the lower density of points in higher dimensional spaces.

\subsection{Extended equivalent layers compression}
Once the profiles are selected, they may need to be compressed to a smaller number of layers in order to be usable in Monte Carlo simulation. \red{Depending on simulation size and complexity, the total number of layers can range from 5 or fewer for SCAO/XAO simulations on small telescopes to over 30 for ELT-scale tomographic AO.} There are several ways of accomplishing this compression that are optimal in different situations \citep{Saxenhuber2017}. For SCAO/XAO, the equivalent layers method can be used \citep{Fusco2001}, which compresses the profile whilst conserving  $r_0$ and $\theta_0$. We present here an extension of the equivalent layers method that also conserves $\tau_0$ by taking into account the wind speed profile. 

The profile is split into $L$ slabs equally spaced in altitude. For each layer $l$ the total turbulence strength is simply integrated 
\begin{equation}
    C_n^2(h_l) \,\mathrm{d}h = \int_l C_n^2(h) \, \mathrm{d}h
\end{equation}
where the integral runs over the altitudes of the slab $l$. The altitude of each layer is given by 
\begin{equation}
    h_l = \frac{\int_l C_n^2(h) h^{5/3} \,\mathrm{d}h}{\int_l C_n^2 (h) \,\mathrm{d}h},
\end{equation}
which ensures conservation of $\theta_0$ by placing the layers at the equivalent altitude. Similarly, we then define the compressed wind velocity profile layers
\begin{equation}
    v(h_l) = \frac{\int_l C_n^2(h) v(h)^{5/3} \, \mathrm{d}h}{\int_l C_n^2 (h) \,\mathrm{d}h},
\end{equation}
which ensures conservation of $\tau_0$. The final required parameter is the wind direction for each compressed layer. We choose to define this through a weighted average of wind directions in each slab, with the weights defined as the wind speeds, i.e.
\begin{equation}
    \theta(h_l) = \mathrm{arg}\left(\frac{\int_l e^{i\theta(h)} v(h) \,\mathrm{d}h}{\int_l v(h) \,\mathrm{d}h}\right)
\end{equation}
where the projection into the complex plane $e^{i\theta}$ ensures correct wrapping of directions when averaging (i.e. $360^\mathrm{o} = 0^\mathrm{o}$). We take the complex argument of the final result to project back into angular units.

\section{Results}
We show an example of this method applied to Stereo-SCIDAR measurements from ESO Paranal \citep{Osborn2018a}, site of the Very Large Telescope and close to the Cerro Armazones site of the Extremely Large Telescope. 

The 2019A dataset consists of 19358 turbulence profiles, \red{taken in 161 nights between April 2016 and October 2019, spanning all months of the year. The average temporal sampling during a run is around two minutes.} Altitude resolution ranges between around 200 m to 1 km, and the profile layers are binned into 250m altitude bins, with 100 bins in total between the ground and 25 km. \red{In addition to turbulence profiles, wind speed and direction for the strongest turbulent layers are also measured. The sparse wind profiles are spatio-temporally interpolated to fill in the missing layers. We note that this may not be an optimal way of obtaining the wind profile, and it may be better to use wind profiles from an external source, for example meteorological models. However, this is beyond the scope of this basic demonstration of the method.}

\begin{figure}
    \centering
    \includegraphics[width=\columnwidth]{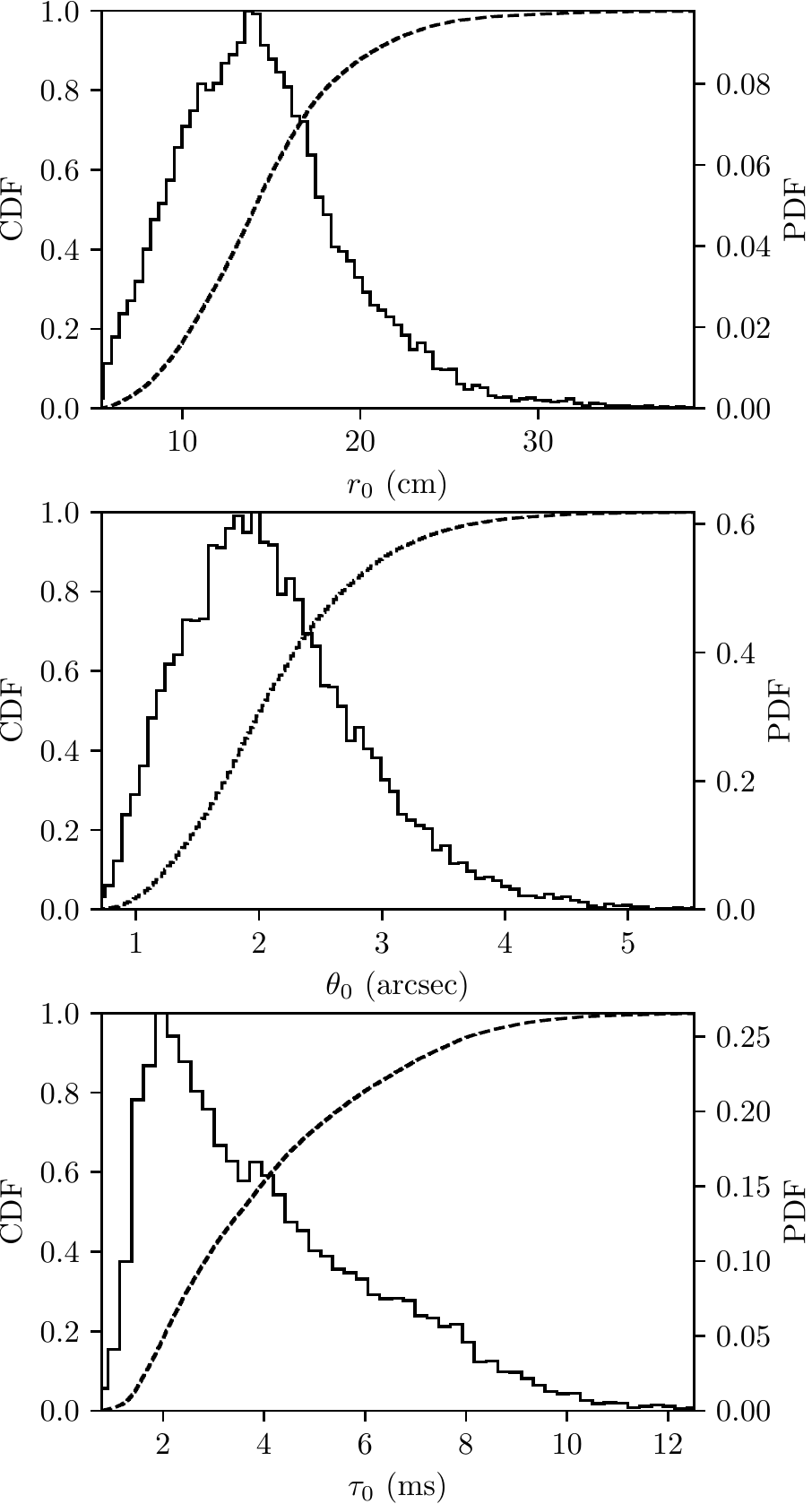}
    \caption{Distributions of $r_0$, $\theta_0$ and $\tau_0$ at $\lambda=500$ nm for the Paranal Stereo-SCIDAR 2019A data release. Each distribution is illustrated by histograms showing probability density (solid lines, right hand axis) and cumulative probability density (dashed lines, left hand axis).}
    \label{fig:params}
\end{figure}

\begin{table}
\centering
\begin{tabular}{ccccc}
\bf{Quantile} & $\boldsymbol{r_0}$ \bf{[cm]} & $\boldsymbol{\theta_0}$ \bf{[arcsec.]} & $\boldsymbol{\tau_0}$ \bf{[ms]} & $\Delta Q$ \\
0.05     & 7.8 (7.8)   & 1.1 (1.1)            & 1.5 (1.5) &  0.003   \\
0.25     & 11.1 (11.0) & 1.6 (1.6)            & 2.3 (2.3) &  0.015  \\
0.5      & 14.1 (14.1) & 2.0 (2.0)            & 3.5 (3.6) &  0.011  \\
0.75     & 17.1 (17.1) & 2.5 (2.6)            & 5.4 (5.3) &  0.018  \\
0.95     & 23.3 (23.0) & 3.5 (3.5)            & 8.3 (8.4) &  0.007  
\end{tabular}
\caption{Quantiles of the distributions of $r_0$, $\theta_0$ and $\tau_0$ from the Paranal Stereo-SCIDAR 2019A dataset. Values in brackets are obtained from the 5 extracted reference profiles, for comparison to the target values. The $\Delta Q$ column indicates the distance in 3-dimensional quantile space between the target and selected profiles.}
\label{tab:params}
\end{table}

The distributions of atmospheric parameters from this dataset are shown in Fig. \ref{fig:params}, with the relevant target quantiles in Tab. \ref{tab:params}. We transform each data point to its quantile within the parameter distributions (Eq. \ref{eq:quantiles}) and employ the k-d tree search algorithm to find profiles that correspond to the target quantiles for each parameter concurrently. The parameters of the resulting profiles are shown in brackets in Tab. \ref{tab:params}, where we see very good agreement with the target values\red{, and $\Delta Q$ values indicating profiles have been found in all cases within a distance 2\% from the target quantiles, and as close as 0.3\%}. 

\begin{figure}
    \centering
    \includegraphics[width=\columnwidth]{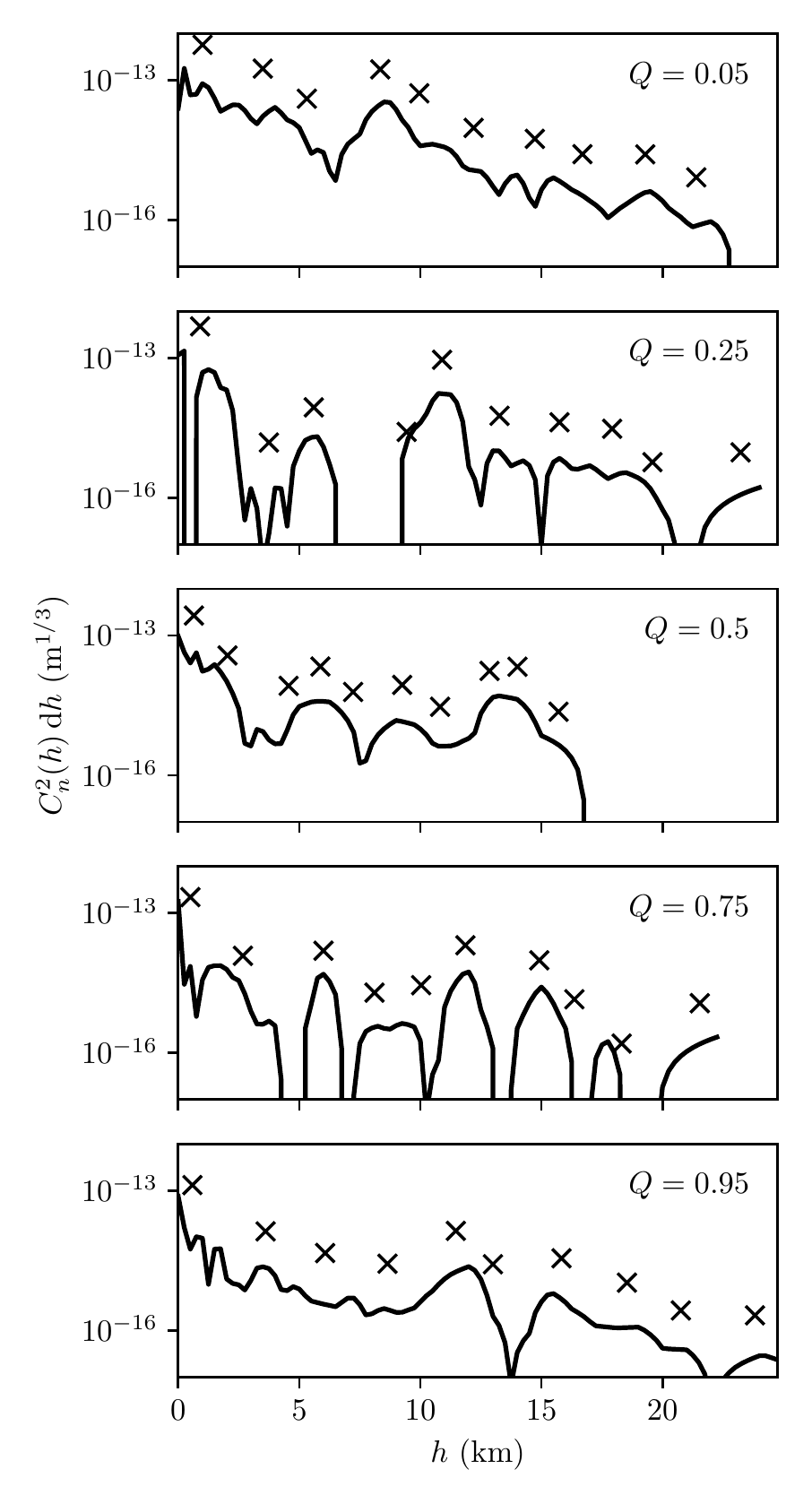}
    \caption{Optical turbulence profiles extracted from the Stereo-SCIDAR 2019A dataset that follow the target quantiles in turbulence parameters. From upper to lower panel: $Q=$0.05, 0.25, 0.5, 0.75 and 0.95, representing very bad conditions to very good conditions. For each panel, the full 100 layer Stereo-SCIDAR is shown by the solid line and a 10 layer compressed version, using our extended equivalent layers method, shown as crosses. When the profile is compressed, the $C_n^2 \, \mathrm{d}h$ values increase since the altitude bins become larger and we are considering the integrated turbulence strength in each altitude bin.}
    \label{fig:cn2}
\end{figure}

\begin{figure}
    \centering
    \includegraphics[width=\columnwidth]{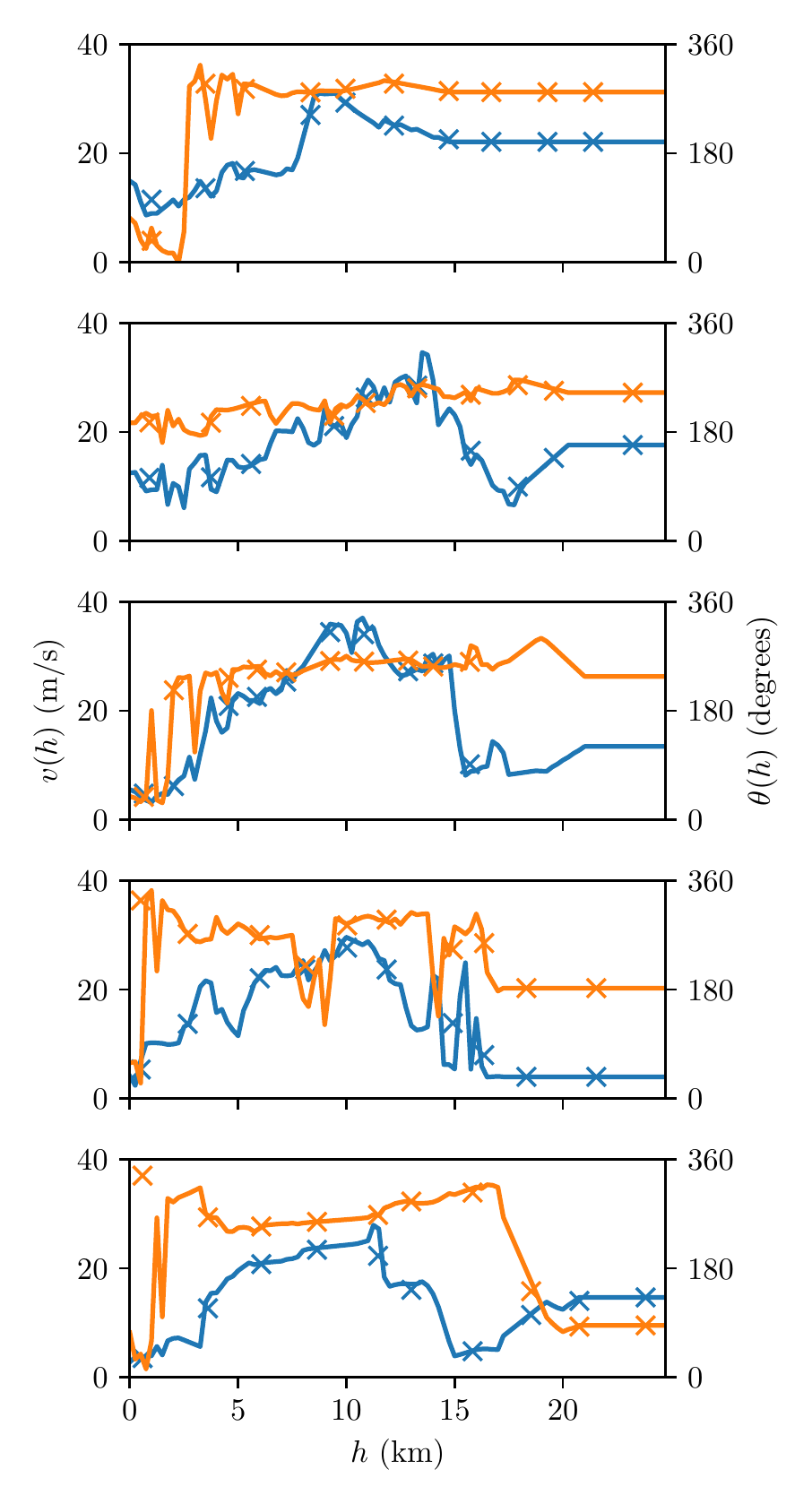}
    \caption{Wind profiles extracted from the Stereo-SCIDAR 2019A dataset that follow the target quantiles in turbulence paramters. From upper to lower panel: $Q=$0.05, 0.25, 0.5, 0.75 and 0.95, representing very bad conditions to very good conditions. Each panel shows both the wind velocity $v(h)$ (blue, left hand axis) and direction $\theta(h)$ (orange, right hand axis) as a function of altitude $h$. The 100 layer Stereo-SCIDAR profiles are shown as solid lines and 10 layer compressed profiles as crosses in each case. }
    \label{fig:wind}
\end{figure}

In Fig. \ref{fig:cn2} we show the full 100 layer profiles and the same profiles compressed to 10 layers with our extended equivalent layers method. The parameters for the compressed profiles are conserved, by definition. We also show wind profiles in Fig. \ref{fig:wind}, with the same compression applied. We can see by simple visual inspection that a diverse range of $C_n^2$ and wind profiles have been extracted.

\begin{figure}
    \centering
    \includegraphics[width=\columnwidth]{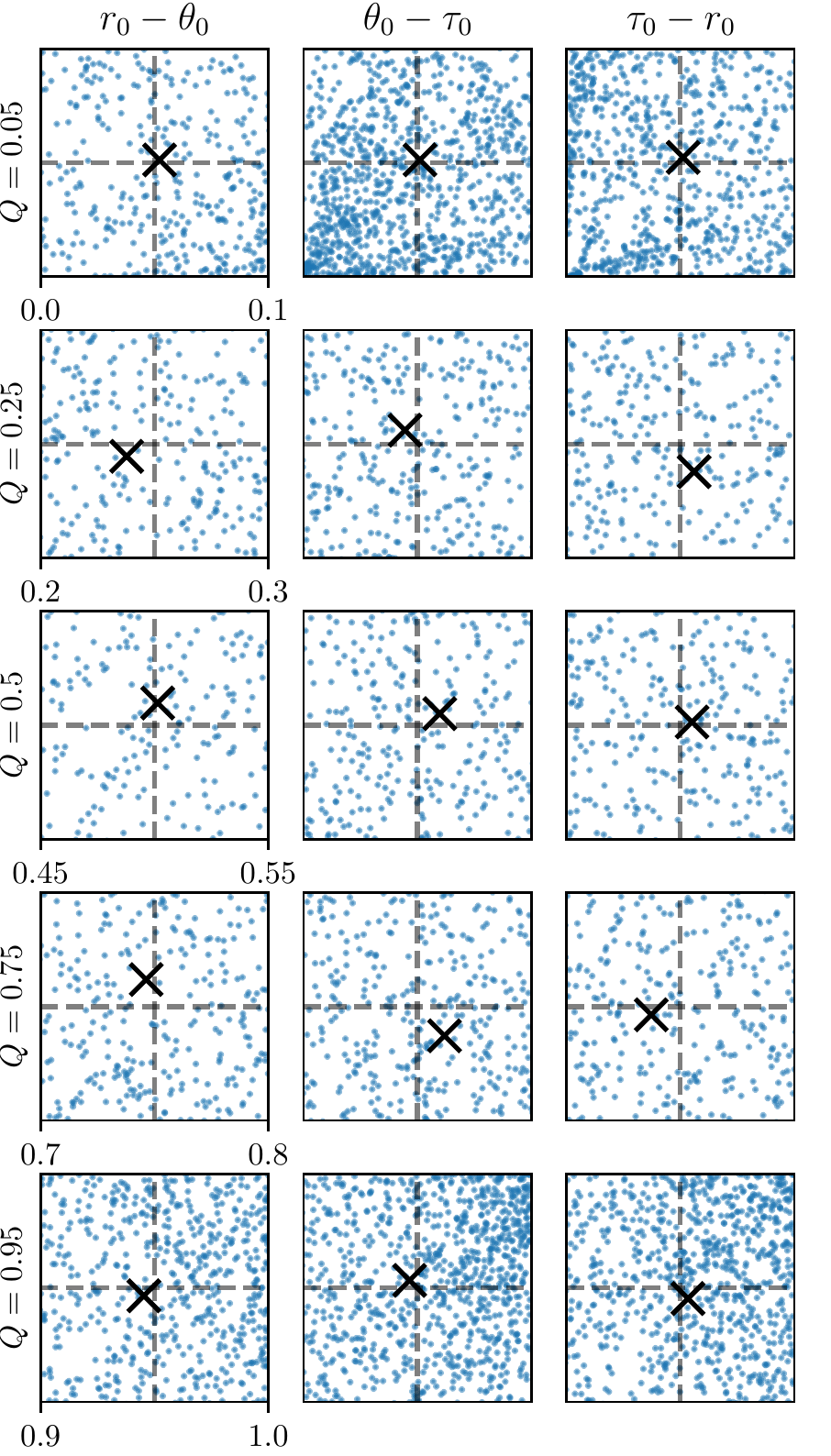}
    \caption{Two dimensional slices through the three dimensional $r_0-\theta_0-\tau_0$ quantile parameter space, showing detail around the target quantiles. Columns represent the $r_0-\theta_0$, $\theta_0-\tau_0$ and $\tau_0-r_0$ planes as indicated at the top of the figure. Each row shows a 0.1 size region around the target quantiles, from $Q=0.05$ to $Q=0.95$ from the upper to lower row. Blue points indicate individual turbulence profiles in the dataset, with the black cross representing the chosen profile for that quantile. Grey dashed lines represent the centre of the target quantile.}
    \label{fig:quantile_slices}
\end{figure}

\red{We illustrate the quality of our selected profiles in Fig. \ref{fig:quantile_slices}, where we show the density of points (profiles) in our 3-dimensional $r_0-\theta_0-\tau_0$ parameter space. We can see that, for the most part, the selected profiles lie very close to their target values. As would be expected from the $\Delta Q$ values in Tab. \ref{tab:params}, this is particularly true for the more extreme ($Q=0.05$ and $Q=0.95$) profiles, where the density of points is greater and so it is more likely a profile can be found close to the target. Points in the parameter space around $Q=0.25,0.5$ and $0.75$ are more sparse, and our selected profile is further from the target, but still within a maximum overall distance of $\Delta Q = 0.018$, or 1.8\% in percentile terms. We highlight the importance of this analysis when employing this method: for smaller datasets or a greater number of parameters, there may not be sufficient density in the parameter space to ensure small $\Delta Q$ values, resulting in profiles that are not representative.}

\begin{figure}
    \centering
    \includegraphics[width=\columnwidth]{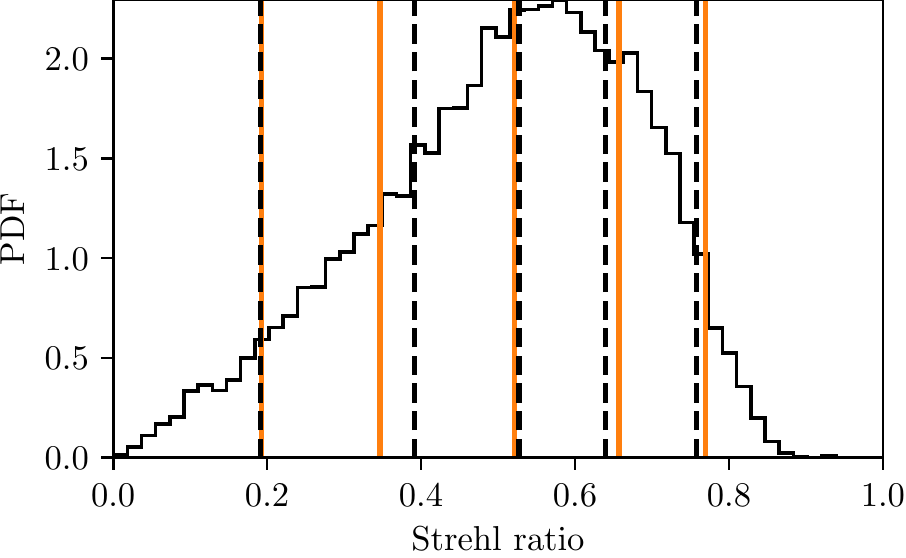}
    \caption{Distribution of Strehl ratio for the 2019A dataset obtained via analytical AO simulation (black histogram), with quantiles $Q=$0.05, 0.25, 0.5, 0.75 and 0.95 indicated by vertical black dashed lines. Strehl ratio obtained with the extracted reference profiles for the same quantiles are indicated by orange vertical lines.}
    \label{fig:strehl}
\end{figure}

Finally, as a form of validation we show results from an analytical Fourier domain AO simulation \citep[][]{Farley2022} in Fig. \ref{fig:strehl}. We compute the on-axis long exposure Strehl ratio for all 19358 profiles, shown as the histogram, and overlay the relevant quantiles of this distribution as black vertical lines. The Strehl ratio obtained with the extracted reference profiles corresponding to the same quantiles are shown as orange vertical lines. The small divergence of the reference profiles from the target quantiles is a result of the fact that for this AO system and simulation, the total wavefront error and hence Strehl ratio is not an exact function of the turbulence parameters as we have assumed. \red{There is also likely a contribution from the fact that some profiles are further from their target values, for example the $Q=0.75$ profile, which corresponds to the second orange line from the left in Fig. \ref{fig:strehl}, has the largest $\Delta Q$ and shows the largest discrepancy of 0.05 in Strehl ratio.} However, for the most part the reference profiles provide very similar Strehl to their target quantiles.

\section{Conclusions}

We have presented a method of extracting reference $C_n^2(h)$, wind velocity and wind direction profiles from large profile datasets for SCAO/XAO. This method relies on the error budgets of these systems being defined by, or at least correlated with, integrated parameters of the turbulence profile.

By computing distributions of these integrated parameters and selecting profiles that adhere to quantiles of these distributions, we obtain a set of profiles that will reflect the variability of the atmosphere when used in simulation. We have also presented an extension of the equivalent layers profile compression method, which may be used to reduce the number of layers in a profile whilst conserving $r_0$, $\theta_0$ and $\tau_0$. 

We have shown the application of this method to the 2019A Stereo-SCIDAR dataset from Paranal, extracting five $C_n^2(h)$ and wind profiles representing a wide range of turbulence conditions.  

Finally we showed that these profiles correspond to a similar range of AO performance, using an analytical AO simulation to compute the long exposure AO-corrected Strehl ratio for every profile in the dataset and our reference profiles. 

\section*{Acknowledgements}

This research made use of Python including NumPy and SciPy \citep{VanderWalt2011}, Matplotlib \citep{Hunter2007}, Astropy, a community-developed core Python package for Astronomy \citep{Robitaille2013} the Python AO utility library AOtools \citep{Townson2019}. 

OJDF acknowledges support from UK Research and Innovation (Future Leaders Fellowship MR/S035338/1). He would like to thank James Osborn for his assistance and Tim Butterley for providing the 2019A Stereo-SCIDAR turbulence profiles. He thanks the anonymous reviewer for their insightful comments and suggestions.

%%%%%%%%%%%%%%%%%%%%%%%%%%%%%%%%%%%%%%%%%%%%%%%%%%
\section*{Data Availability}

The data underlying this article will be shared on reasonable request to the corresponding author.



\bibliographystyle{mnras}
\bibliography{references1} % if your bibtex file is called example.bib

%\begin{thebibliography}{99}
%\end{thebibliography}

%%%%%%%%%%%%%%%%%%%%%%%%%%%%%%%%%%%%%%%%%%%%%%%%%%

%%%%%%%%%%%%%%%%% APPENDICES %%%%%%%%%%%%%%%%%%%%%

%%%%%%%%%%%%%%%%%%%%%%%%%%%%%%%%%%%%%%%%%%%%%%%%%%

% Don't change these lines
\bsp	% typesetting comment
\label{lastpage}
\end{document}